\documentclass[showpacs,twocolumn,preprintnumbers,amsmath,amssymb]{revtex4}
\usepackage{mathrsfs}
\usepackage{graphicx}
\usepackage{subfigure}
\usepackage{amsfonts}
\usepackage{amssymb}%
\usepackage{indentfirst}
\usepackage{color}

\def\ga{\gamma}
\def\de{\delta}

\def\ga{\gamma}

\def\ka{\kappa}
\def\la{\lambda}

\def\si{\sigma}
\def\Ga{\Gamma}

\def\prt{\partial}

\def\mn{{\mu\nu}}

\newcommand{\bea}{\begin{eqnarray}}
\newcommand{\eea}{\end{eqnarray}}
\newcommand{\bit}{\begin{itemize}}
\newcommand{\eit}{\end{itemize}}

\newcommand{\hf}{\frac{1}{2}}
\newcommand{\nn}{\nonumber\\}
\newcommand{\ie}{{\it i.e.}}
\newcommand{\etc}{{\it etc}}
\newcommand{\eg}{{\it e.g.}}
\newcommand{\etl}{{\it et.al.}}

\providecommand{\Journal}[4] {#1 {\bf#2}, #4 (#3)}
\providecommand{\AoM}{Ann. of Math.} %
\providecommand{\AJP}{Am. J. Phys.} %
\providecommand{\ASR}{Adv. Space Res.} %
\providecommand{\IJMPD}{Int. J. Mod. Phys. D} %
\providecommand{\CQG}{Class. Quantum Grav.}
\providecommand{\EJPS}{Euro Jnl. Phil. Sci.}
\providecommand{\LRR}{Living Rev. Relativity}
\providecommand{\PR}{Phys. Rev.} %
\providecommand{\PRt}{Phys. Rept.} %
\providecommand{\PRL}{Phys. Rev. Lett.} %
\providecommand{\PRD}{Phys. Rev. D.} %
\providecommand{\PLA}{Phys. Lett. A} %
\providecommand{\PLB}{Phys. Lett. B} %
\providecommand{\PL}{Phys. Lett.} %
\providecommand{\NT}{Nature} %
\providecommand{\NP}{Nature Phys.} %
\providecommand{\NC}{Nat. Commun.} %
\providecommand{\NPB}{Nucl. Phys. B} %
\providecommand{\AP}{Ann.Phys.} %
\providecommand{\JMP}{J. Math. Phys.} %
\providecommand{\JHEP}{JHEP} %
\providecommand{\JPCS}{J. Phys. Conf. Ser.} %
\providecommand{\RPP}{Rep. Prog. Phys.} %
\providecommand{\GRG}{Gen. Relativ. Gravit.} %
\providecommand{\RMP}{Rev. Mod. Phys.} %
\providecommand{\SCPMA}{Sci. China Phys. Mech. Astron.} %
\providecommand{\Sym}{Symmetry}

\begin{document}
\title{Lorentz-violating scalar Hamiltonian and equivalence principle in a static metric}
\author{Zhi Xiao}
\email{blueseacat@126.com}
\affiliation{Department of Mathematics and Physics, North China Electric Power University, Beijing 102206, China}%\\
%$^2$Indiana University Center for Spacetime Symmetries (IUCSS)}

\begin{abstract}
In this paper, we obtain a nonrelativistic Hamiltonian from the Lorentz-violating (LV)
scalar Lagrangian in the minimal SME.
The Hamiltonian is obtained by two different methods. One is through the usual ansatz
$\Phi(t,\vec{r})=e^{-imt}\Psi(t,\vec{r})$ applied to the LV corrected Klein-Gordon equation, and the other
is the Foldy-Wouthuysen transformation.
The consistency of our results is also partially supported by the comparison with the spin-independent part of the fermion Hamiltonian.
In this comparison, we can also establish a relation between
the set of scalar LV coefficients with their fermion counterparts.
Using a pedagogical definition of the weak equivalence principle (WEP), we further point out that the LV Hamiltonian
not only necessarily violates universal free fall,
which is clearly demonstrated in the geodesic deviation, but also violates WEP in a semi-classical setting.
As a bosonic complement, this method can be straightforwardly applicable to the spin-1 case, which
shall be useful in the analysis of atomic tests of WEP, such as the case of the $^{87}\text{Rb}_1$ atom.
\end{abstract}
\maketitle

\section{Introduction}
Symmetry has been a main theme of physics in the last century, and may continue to be so in 21st century. Of the various kind of symmetries we know,
Local Lorentz symmetry (LLS) is the most fundamental. It is a cornerstone of the Standard Model (SM) in particle physics and General Relativity (GR).
Though SM and GR have achieved impressive successes with various experimental verifications \cite{HiggsLHC}\cite{GravWave},
there is still no concrete clue on a consistent theory of quantum gravity (QG),
which may help to resolve longstanding puzzles in contemporary physics, such as the intriguing information paradox inside black hole \cite{FireWall}.
On the other hand, there is a growing interest in searching for tiny violations of Lorentz symmetry
both in theory \cite{Update} and experiment \cite{DATA}.
Indeed, many candidate QG theories predict such a possibility \cite{PLV}. If proved to be true, it will definitely be
a concrete clue to the physics at Planck scale, an ultrahigh energy scale far beyond any direct experimental access.
To thoroughly explore this possibility, Kosteleck\'y and his collaborators have established an effective field theory called
Standard Model Extension (SME) \cite{sme}\cite{2004Alan}\cite{Nonmini}, which incorporates SM and GR, with various possible LV operators. This framework largely facilitates the study of Lorentz and CPT symmetry,
and has already become a powerful toolbox in both theoretical and phenomenological investigations in this field \cite{CPTM}.

As another conceptual bridge from special relativity to GR, the equivalence principle (EP),
especially the Einstein equivalence principle (EEP), entails a close relationship to Lorentz symmetry
and has also been broadly tested in
various kinds of physical systems \cite{EPTheo}\cite{EPExp}\cite{TorsBan2012}\cite{Shao2016}.
According to the famous statement by C.M. Will \cite{EPCW}, LLS, local position invariance and the weak equivalence principle (WEP) are the three key ingredients of EEP.
So violation of LLS necessarily implies violation of EEP, while the contrary is not necessarily true.
A thorough investigation of the relation between EP and LLS is still missing \cite{SWLSEP}\cite{MapEPLV}, though in view of Schiff's conjecture \cite{Schiff},
WEP may imply the validity of LLS.
Moreover, even in the Lorentz-invariant (LI) context, the debate as to whether EP holds true in the quantum domain
seems still far from closing \cite{Ahluwalia}\cite{OkonEP}.
In this paper, we do not try to involve too much into this debate.
Instead, we adopt a relatively conservative point of view,
\ie, there is no conflict of WEP with nonrelativistic (NR) quantum mechanics \cite{OkonEP}\cite{EPOK}. In other words,
the NR Hamiltonian derived from GR for the Schr$\ddot{\mathrm{o}}$dinger equation is compatible with WEP.
For any nonrelativistic system, the well-known Bargmann's super-selection rule prohibits mass from being a superposition parameter \cite{Barg},
thus superposition of different mass eigenstates, like neutrino oscillation in relativistic physics, is beyond the scope of this constrained assertion.
Taking into account the fact \cite{TorsBan2012}\cite{MicroscopeM}\cite{SpinWPT} that most laboratory tests up to now have still being nonrelativistic,
we think an appropriate test framework for WEP even in the quantum regime must go beyond GR (test of WEP in the classical domain necessarily go beyond GR).

Many generalized theories of gravity \cite{EPTheo}\cite{EPCW}\cite{FTGKip}\cite{WTNEP} fit into this category,
but in our viewpoint, the gravity sector of SME \cite{2004Alan} is more suitable for such a task.
In SME, WEP violation is associated with Lorentz and CPT violation since various LV coefficients can also be species-dependent,
which enables more exotic violation effects \cite{MattGC} and makes this framework as broad as it can be.
Discussions of EP in this framework are also abundant \cite{MattGC}\cite{YuriEP}\cite{YuriEPI}\cite{EXPEPL}, and
most of them concentrate on fermion-gravity couplings, since matter is composed of fermions.
However, in an effective point of view, as the test particles can also be composite bosons made of fermions, such as $^{88}$Sr or $^{133}$Cs,
we think it would be a valuable complementary to discuss EP directly using boson fields instead,
especially taking account of the recent trend in utilizing microscopic objects such as cold atoms
as test particles \cite{EPTheo}\cite{SpinWPT}\cite{EntangA}.
In this sense, the boson LV coefficients can be totally effective, \ie, microscopically,
they must be certain combinations of the LV coefficients of the component fermions involved
(\eg, electron and proton).
In this paper, for simplicity, we focus on the scalar.

The paper is organized as follows. In the next section, we briefly review the scalar LV Lagrangian and the corresponding canonical formalism.
In Sec. \ref{TNRKG}, by using the ansatz $\Phi(t,\vec{r})=e^{-imt}\psi(t,\vec{r})$, we derive the NR Hamiltonian to first order in LV coefficients
and metric perturbations from the LV corrected Klein-Gordon equation.
In Sec. \ref{SFWT}, following the method of \cite{CVZRW}, we recast the Klein-Gordon equation into the Schr$\ddot{\mathrm{o}}$dinger form,
then to the desired order of approximation, we get the NR Hamiltonian using the Foldy-Wouthuysen transformation (FWT) \cite{FW}\cite{Case}. In Sec. \ref{EPSNR},
we briefly discuss the test of EP and its possible relevance to the Hamiltonian we derived.
Then we summarize our results in section \ref{Summary}.
The convention is the same as in \cite{2004Alan}, where $\mathrm{diag}(\eta_\mn)=(-1,1,1,1)$ and $\epsilon_{0123}=+1$.

\section{Hamiltonian of the Lorentz-Violating Scalar}
In \cite{2004Alan}, by generalizing SME to Riemann-Cartan spacetime,
Kosteleck\'y introduced various LV operators both in the pure gravity sector and in the matter sector
through minimal matter-gravity couplings.
In the matter sector, the Higgs Lagrangian reads
\bea\label{LCLVS}&&
\mathcal{L}_{\Phi}=-e\left\{[g^{\mn}-(\tilde{k}_{\phi\phi})^{\mn}]D_\mu\Phi^\dagger{D}_\nu\Phi
+(m^2+\xi{R})\Phi^\dagger\Phi\right.\nn&&
~~~\left.-[i(k_\phi)^\mu\Phi^\dagger{D}_\mu\Phi+h.c.]+\hf{k_{\phi{A}}}^{\mn}F_{\mn}\Phi^\dagger\Phi\right\},
\eea
where $D_\nu\Phi=(\nabla_\nu-iqA_\nu)\Phi$, and for completeness,
we also included the non-minmal coupling $\xi{R}$ term.
Note that for notational simplicity, we have introduced $(\tilde{k}_{\phi\phi})^{\mn}\equiv\hf[(k_{\phi\phi})^{\mn}+{(k_{\phi\phi})^{\nu\mu}}^*]$,
which can be taken to have  a symmetric real part and an antisymmetric imaginary part. $(k_\phi)^\mu$ can also take complex values,
though in flat spacetime it must be real.
For later convenience, we can further define
$\tilde{k}_{\phi\phi}^{\mn}\equiv(K^{\mn}+iS^{\mn})$ with $K^{\mn}=K^{\nu\mu},S^{\mn}=-S^{\nu\mu}$,
and $K^{\mn},S^{\mn}\in\mathbb{R}$.
Similarly, we can also define $(k_\phi)^\mu\equiv(a^\mu+ib^\mu)$ with $a^\mu,b^\mu\in\mathbb{R}$.
As mentioned before, here $(k_\phi)^\mu, \tilde{k}_{\phi\phi}^{\mn}$ can be regarded as effective LV
coefficients of composite spin-0 bosons, not necessarily referring to the LV coefficients of the Higgs particle.

From the Lagrangian (\ref{LCLVS}), we can define $\tilde{G}^{\mn}\equiv[g^{\mn}-(\tilde{k}_{\phi\phi})^{\mn}]$.
Then the Euler-Lagrangian equation is given by
\bea\label{SEOM}&&
[D_\mu+\frac{\prt_\mu{e}}{e}]\left[\tilde{G}^{\mn}D_\nu\Phi+i{k_\phi^{~\mu}}^*\Phi\right]+i{k_\phi}^\mu{D}_\mu\Phi\nn&&
~~~-\hf{k_{\phi{A}}}^{\mn}F_{\mn}\Phi-(m^2+\xi{R})\Phi
=0.
\eea
This equation is intrinsically second-order in time derivatives, so we cannot obtain
a Schr$\ddot{\mathrm{o}}$dinger-like equation directly from (\ref{SEOM}). Instead, we
turn to the canonical formalism.
From
{\small
\bea\label{CanonM1}&&
\pi_{\Phi}\equiv\frac{\prt\mathcal{L}_{\Phi}}{\prt\dot{\Phi}}=
-e\left[\tilde{G}^{\rho0}(D_\rho\Phi)^\dagger-ik_\phi^{~0}\Phi^\dagger\right],\\&&\label{CanonM2}
\pi_{\Phi^\dagger}\equiv\frac{\prt\mathcal{L}_{\Phi}}{\prt\dot{\Phi}^\dagger}=
-e\left[\tilde{G}^{0\rho}D_\rho\Phi+i{k_\phi^{~0}}^*\Phi\right],
\eea
}
we can solve $\dot{\Phi},~\dot{\Phi}^\dagger$ in terms of $\pi_{\Phi},~\pi_{\Phi^\dagger}$, \ie,
{\small
\bea\label{CanonK1}&&
\dot{\Phi}^\dagger=\frac{-1}{\tilde{G}^{00}}\left[\frac{\pi_\Phi}{e}-ik_\phi^{~0}\Phi^\dagger+\tilde{G}^{i0}(D_i\Phi)^\dagger\right]-iqA_0\Phi^\dagger,\\&&\label{CanonK2}
\dot{\Phi}=\frac{-1}{\tilde{G}^{00}}\left[\frac{\pi_{\Phi^\dagger}}{e}+i{k_\phi^{~0}}^*\Phi
+\tilde{G}^{0i}D_i\Phi\right]+iqA_0\Phi.
\eea}
Performing the canonical transformation on (\ref{LCLVS}), we get the Hamiltonian density
\bea\label{CompHamD}&&
\mathcal{H}=-\frac{\pi_\Phi\pi_{\Phi^\dagger}}{e\tilde{G}^{00}}-\frac{1}{\tilde{G}^{00}}
\left[\tilde{G}^{0j}\pi_\Phi{D}_j\Phi+\tilde{G}^{j0}(D_j\Phi)^\dagger\pi_{\Phi^\dagger}\right]
\nn&&
~~~+i\left[\frac{k^0_\phi}{\tilde{G}^{00}}-qA_0\right]\Phi^\dagger\pi_{\Phi^\dagger}
-i\left[\frac{{k^0_\phi}^*}{\tilde{G}^{00}}-qA_0\right]\pi_\Phi\Phi
\nn&&
~~~+e\left[\bar{G}^{ij}(D_i\Phi)^\dagger{D_j}\Phi+\bar{M}^2\Phi^\dagger\Phi\right]
+ie\left\{[{k_\phi^j}^*(D_j\Phi)^\dagger\Phi\right.\nn&&
~~~\left.-k_\phi^j\Phi^\dagger{D_j}\Phi]+\frac{1}{\tilde{G}^{00}}[k^0_\phi\Phi^\dagger\tilde{G}^{0j}{D_j}\Phi\right.
\nn&&~~~
\left.-{k^0_\phi}^*\tilde{G}^{j0}(D_j\Phi)^\dagger\Phi]\right\},
\eea
where we have defined $\bar{G}^{ij}\equiv[\tilde{G}^{ij}-\frac{\tilde{G}^{i0}\tilde{G}^{0j}}{\tilde{G}^{00}}]$ and
$\bar{M}^2\equiv\left[m^2+\xi{R}+\hf{k_{A\phi}}\cdot{F}-\frac{|k^0_\phi|^2}{\tilde{G}^{00}}\right]$. Also
note ${\bar{G}^{ij}}{^*}=\tilde{G}^{ji}-\frac{\tilde{G}^{0i}\tilde{G}^{j0}}{\tilde{G}^{00}}=\bar{G}^{ji}$,
since ${(\tilde{k}_{\phi\phi})^{\mn}}^*=(\tilde{k}_{\phi\phi})^{\nu\mu}$.
The Hamiltonian density (\ref{CompHamD}) will be useful in section \ref{SFWT} for the derivation of a Schr$\ddot{\mathrm{o}}$dinger-like equation.
Before the end of this section, we mention that we will set $A_\mu=0$ to
avoid electromagnetic interaction in the following sections,
as even a very tiny electromagnetic interaction spoils the test of WEP,
and we included it here only for completeness. Strictly speaking, only neutral particle is immune to electromagnetic interaction,
and in that case the scalar field must be real. In flat spacetime, we can discard $(k_\phi)^\mu$ term as
it only contributes a total derivative for a real scalar.
Similarly, $(\tilde{k}_{\phi\phi})^{\mn}$ can only take the real symmetric and traceless part,
and can be shifted to the fermion sector with $c_\mn\rightarrow{c_\mn-\hf(\tilde{k}_{\phi\phi})^{\mn}}$
through coordinate transformation \cite{MattGC}.
%$x^\mu\rightarrow{x^{\mu'}}=x^\mu-\hf{(\tilde{k}_{\phi\phi})^\mu_{~\rho}}x^\rho$
However, all the above issues are not very relevant here when coupled with gravity.
For a gravity coupled neutral scalar, we only need to ignore the $S^{\mn}$ and $k_{\phi{A}}^{\mn}$ terms.
For completeness, below we will still use the complex scalar to demonstrate all the results.

\section{Static Metric and Traditional Route to the Non-relativistic Equation}\label{TNRKG}
In curved spacetime, LV coefficients can also contribute to the energy momentum tensor \cite{2004Alan}
and, through the Einstein equation, affect the corresponding metric solutions. Here, since the statement of WEP involves a ``free-moving" test particle
and we are only interested in matter-gravity couplings, for simplicity we can adopt a test particle assumption \cite{YuriEPI},
where spacetime metric is untouched by the LV coefficients associated with the matter sector. So we
can still make use of the conventional metric from GR, and ``free motion'' implies we have to take $A_\mu=0$ in (\ref{SEOM}), which gives
\bea\label{SEOM2}&&
\left\{g^{\mn}\nabla_\mu\nabla_\nu-\tilde{k}_{\phi\phi}^{\mn}[\prt_\mu\prt_\nu+\Ga^\la_{~\mu\la}\prt_\nu]
+i{k_\phi^\mu}^*[\prt_\mu+\Ga^\la_{~\mu\la}]
\right.\nn&&~~~\left.+i{k_\phi}^\mu\prt_\mu-(m^2+\xi{R})\right\}\Phi=0,
\eea
where for simplicity we also assumed Riemann spacetime instead of Riemann-Cartan spacetime,
otherwise $\frac{1}{e}\prt_\mu[eg^{\mn}\prt_\nu]\Phi=g^{\mn}[\nabla_\mu\nabla_\nu-T_{(\mn)}^{~~~\la}\nabla_\la]\Phi$,
where $T^\la_{~\mn}$ is the torsion tensor.
For simplicity, we can take the isotropic static metric \cite{SGI}
\bea\label{SMetric}
ds^2=-g_{\mn}dx^\mu{dx}^\nu=V^2dt^2-\de_{\hat{i}\hat{j}}W^2dx^idx^j
\eea
as an example.
Then the only non-zero Christoffel symbols are given by
\bea\label{AffineC}&&
\Ga^i_{~jk}=[\de^i_j\prt_kW+\de^i_k\prt_jW-\de_{jk}\prt_iW]/W,\quad \Ga^0_{~0j}=\frac{\prt_jV}{V},\nn&&%\quad
\Ga^j_{~00}=\hf\frac{\prt_jV^2}{W^2},\quad \Ga^\la_{i\la}=\prt_iV/V+3\prt_iW/W.
\eea
\begin{widetext}
Defining $\mathcal{F}\equiv\frac{V}{W}$, and substituting (\ref{AffineC}), $g^{00}=-1/V^2$ and $g^{ij}=\delta^{ij}/W^2$ into (\ref{SEOM2}), we get
{\small
\bea\label{LVSE2}&&
\left\{-\prt_0^2+\mathcal{F}^2\left[\Delta+\vec{\nabla}\ln(VW)\cdot\vec{\nabla}\right]
-V^2(m^2+\xi{R})\right\}\Phi%\nn&&~~~
=V^2\left\{\tilde{k}_{\phi\phi}^{\mn}\left[\prt_\mu\prt_\nu
+\delta^i_\mu\prt_i\ln(VW^3)\prt_\nu\right]%\right.\nn&&~~~~~~~~~~
%\left.
-\left[2ia^\mu\prt_\mu+i{k_\phi^j}^*\prt_j\ln(VW^3)\right]\right\}\Phi.\nn
\eea
}
The Ricci scalar for the metric (\ref{SMetric}) is given by
\bea&&
R=\frac{2}{VW^4}\left[W^2\nabla^2V+2WV\nabla^2W+W\vec{\nabla}V\cdot\vec{\nabla}W%\right.\nn&&~~~\left.
-V(\vec{\nabla}W)^2\right],
\eea
Note $R$ differs by a minus sign if using convention $\mathrm{diag}(\eta_\mn)=(1,-1,-1,-1)$.\\

Now substituting the ansatz $\Phi(t,\vec{r})=e^{-imt}\psi(t,\vec{r})$ into (\ref{LVSE2}), we can get
\iffalse
\bea\label{TediEOM}&&
[m^2\psi+2im\dot{\psi}-\ddot{\psi}]
+\frac{\mathcal{F}^2}{1+\tilde{k}_{\phi\phi}^{00}V^2}[\vec{\nabla}^2+\vec{\nabla}\ln(VW)\cdot\vec{\nabla}]\psi+\frac{V^2}{1+\tilde{k}_{\phi\phi}^{00}V^2}\left\{-(m^2+\xi{R})
+\tilde{k}_{\phi\phi}^{(0i)}(im\prt_i-\prt_i\prt_0)-\tilde{k}_{\phi\phi}^{ij}\prt_i\prt_j\right.\nn&&~~~
\left.
+\nabla_i\ln(VW^3)[\tilde{k}_{\phi\phi}^{i0}(im-\prt_0)-\tilde{k}_{\phi\phi}^{ij}\prt_j]+i\left[2a^0(\prt_0-im)+2\vec{a}\cdot\vec{\nabla}
+(\vec{a}-i\vec{b})\cdot\vec{\nabla}\ln(VW^3)\right]\right\}\psi=0,
\eea
\fi
\bea\label{TediEOM}&&
\left[m^2\psi+2im\dot{\psi}-\ddot{\psi}\right]
+\frac{\mathcal{F}^2}{1+\tilde{k}_{\phi\phi}^{00}V^2}\left[\Delta+\vec{\nabla}\ln(VW)\cdot\vec{\nabla}
-W^2(m^2+\xi{R})\right]\psi=
\frac{V^2}{1+\tilde{k}_{\phi\phi}^{00}V^2}\left\{
\left[\tilde{k}_{\phi\phi}^{i0}\prt_i\ln(VW^3)
\right.\right.\nn&&~~
\left.+\tilde{k}_{\phi\phi}^{(0i)}\prt_i\right]
(\prt_0-im)+\tilde{k}_{\phi\phi}^{ij}\left[\prt_i+\prt_i\ln(VW^3)\right]\prt_j
\left.
-i\left[2a^0(\prt_0-im)+2\vec{a}\cdot\vec{\nabla}
+(\vec{a}-i\vec{b})\cdot\vec{\nabla}\ln(VW^3)\right]\right\}\psi,
\eea
where $\tilde{k}_{\phi\phi}^{(0i)}\equiv(\tilde{k}_{\phi\phi}^{0i}+\tilde{k}_{\phi\phi}^{i0})$.
Since most of the tests of EP and LLS up-to-date have been done near the Earth's surface,
where the metric functions are asymptotically flat, \ie, $g_{\mn}\simeq\eta_{\mn}$,
we can resort to the approximation scheme in \cite{MattGC},
where terms proportional to the product of LV coefficients and metric perturbation of powers of $l$ and $n$ respectively are denoted by $\mathcal{O}(l,n)$.
Next, we proceed our calculations with the Schwarzschild metric
$V=(1+\hf\chi)(1-\hf\chi)^{-1},~W=(1-\hf\chi)^2$, where $\chi\equiv-\frac{GM}{c^2r}$.
Now $R=0$, when $r\neq0$, even $R^\ka_{~\la\mn}\neq0$ in general.
Below, we will expand $g_{\mn}$ in powers of $\chi$, and keep only terms up
to $\mathcal{O}(0,2)$ and $\mathcal{O}(1,1)$. In doing so, we also take advantage of the
Virial theorem that $\chi\sim\frac{\bar{v}^2}{c^2}$. In essence, that means we can also take $\bar{v}$ (assuming in natural units that $c=1$) as an expansion parameter.
Also note that in laboratory experiments, $|\prt_i\chi|\ll|\chi/L|$ \cite{MattGC},
where $L$ is the typical experimental scale,
so we can treat $\nabla_i\chi$ as higher order compared to $\chi$, and ignore its product with LV coefficients.
Under these assumptions, we can rearrange (\ref{TediEOM}) as below,
%intermediate step
\bea\label{IntSt1}&&
i\dot{\psi}=\left\{\mathcal{F}^2\left[1-(\tilde{k}_{\phi\phi}^{00}+\frac{a^0}{m})V^2\right]\frac{\hat{\vec{p}}^2}{2m}
-\frac{\mathcal{F}^2}{2m}\vec{\nabla}\ln(VW)\cdot\vec{\nabla}+\frac{m}{2}\left[(V^2-1)-\tilde{k}_{\phi\phi}^{00}V^4\right]\right.
\nn&&~~~\left.
+V^2\left[\frac{\tilde{k}_{\phi\phi}^{(0i)}}{2}\hat{p}_i-\frac{\tilde{k}_{\phi\phi}^{ij}}{2m}\hat{p}_i\hat{p}_j
+\frac{\vec{a}\cdot\hat{\vec{p}}}{m}-\frac{a^0}{2}(V^2+1)\right]
\right\}\psi+V^2\frac{\tilde{k}_{\phi\phi}^{(0i)}}{2m}\nabla_i\dot{\psi}
+\left(1-\frac{a^0}{m}V^2\right)\frac{\ddot{\psi}}{2m}.
\eea
At order $\mathcal{O}(0,1)$, we have $i\dot{\psi}=[-\frac{\vec{\nabla}^2}{2m}+m\chi]\psi$, which is roughly the order of $m\bar{v}^2$. So we know $\frac{\ddot{\psi}}{2m}\sim{m}(\bar{v}^2)^2\sim{m\chi^2}$, and then we can temporarily ignore the last two terms proportional to $\nabla_i\dot{\psi}$ and $\frac{\ddot{\psi}}{2m}$ in (\ref{IntSt1}), and get
\bea\label{TemEOM}&&
i\dot{\psi}=\left\{\left[(1+4\chi)-(\tilde{k}_{\phi\phi}^{00}+\frac{a^0}{m})(1+6\chi)\right]\frac{\hat{\vec{p}}^2}{2m}
+\frac{\chi}{4m}\vec{\nabla}\chi\cdot\vec{\nabla}
+\left[m\chi(1+\chi)-m\tilde{k}_{\phi\phi}^{00}(\hf+2\chi)\right]-(1+3\chi)a^0\right.\nn&&~~~~
+\left.(1+2\chi)\left[\frac{\tilde{k}_{\phi\phi}^{(0i)}}{2}\hat{p}_i-\frac{\tilde{k}_{\phi\phi}^{ij}}{2m}\hat{p}_i\hat{p}_j
+\frac{\vec{a}\cdot\hat{\vec{p}}}{m}\right]\right\}\psi
\eea
up to $\mathcal{O}(\chi^2)$, except for the LI term $\frac{\chi}{4m}\vec{\nabla}\chi\cdot\vec{\nabla}$.
Now defining the terms in the large braces in (\ref{TemEOM}) as $\hat{H}_0$,
and adding the correction $\frac{\ddot{\psi}}{2m}=-\frac{1}{2m}(\hat{H}_0)^2\psi$ and $\nabla_i\dot{\psi}=-i\nabla_i(\hat{H}_0\psi)$
back into (\ref{TemEOM}) to replace the last two terms in (\ref{IntSt1}), we can get up to the desired order,
{\small
\bea\label{2ndAPP}&&
i\dot{\psi}=\left\{\left[(1+3\chi-\frac{\tilde{k}_{\phi\phi}^{00}}{2})\frac{\hat{\vec{p}}^2}{2m}+m\chi(1+\frac{\chi}{2})
-\frac{(\hat{\vec{p}}^2)^2}{8m^3}\right]
+(1+\chi)\left[\frac{\vec{a}\cdot{\hat{\vec{p}}}}{m}-\frac{\tilde{k}_{\phi\phi}^{ij}}{2m}\hat{p}_i\hat{p}_j\right]
+(1+2\chi)\left[\frac{\tilde{k}_{\phi\phi}^{(0i)}}{2}\hat{p}_i-a^0\right]\right.\nn&&~~
-\frac{m}{2}\tilde{k}_{\phi\phi}^{00}(1+3\chi){\Bigg\}}\psi+
\left\{\frac{i}{2m}(1+\frac{13\chi}{2})\vec{\nabla}\chi\cdot\hat{\vec{p}}+
\left[2\frac{a^0}{m}-\tilde{k}_{\phi\phi}^{00}\right]\chi
\frac{\hat{\vec{p}}^2}{2m}
+\frac{1}{4m}(\Delta\chi+2(\vec{\nabla}\chi)^2)
+\tilde{k}_{\phi\phi}^{(0i)}\frac{\hat{\vec{p}}^2}{4m^2}\hat{p}_i\right\}\psi.
\eea
}
\end{widetext}
Note that, as the procedure implies, the above equation will be valid only up to
$\mathcal{O}(0,2)$ and $\mathcal{O}(1,1)$. We divide the right hand side of
(\ref{2ndAPP}) into two parts.
In fact, comparing with the NR Hamiltonian (\ref{NonRH})
obtained by a quite different method,
we find that except for the $\vec{\nabla}\chi\cdot\hat{\vec{p}}$ term (belonging to the
latter brace), the part enclosed by the former brace is consistent with (\ref{NonRH}) up
to the desired orders, while those in the latter brace may be classified as divergent higher order terms.
Indeed we can even verify this coincidence (of the NR results obtained with different methods)
by choosing another metric, for example, the uniform accelerating metric.
So it is interesting to explore whether the above NR procedure can be improved to yield completely
consistent results with the FWT, or even extended to higher orders systematically.
This question is beyond the scope of this paper. In the next section, we will utilize
the FWT \cite{CVZRW}\cite{FW}\cite{Case}\cite{FVWMP}\cite{ABFW}
to show that, the NR approximation can indeed be obtained systematically.

\section{Schr$\ddot{\mathrm{O}}$dinger-like Equation for Scalar field and FWT}\label{SFWT}
The Foldy-Wouthuysen transformation for scalar field was first introduced in \cite{Case}, and later refined by \cite{CVZRW}\cite{FVWMP}\cite{ABFW}. In order to perform FWT for scalar field, first we have to obtain a Schr$\ddot{\mathrm{o}}$dinger-like Hamiltonian from the scalar Lagrangian (\ref{LCLVS}), then we can do
a pseudounitary transformation parallel to the case of the fermion, then with a series expansion in terms of $\frac{1}{m}$,
we can obtain the NR approximation to any desired order we like. Below, we will show the FWT up to $\mathcal{O}(1,1),~\mathcal{O}(0,2)$ in a static Schwarzschild metric,
and we will perform the FWT both directly \cite{Case}\cite{FVWMP} and
indirectly with a unitary transformation \cite{CVZRW} performed first.
We will show that these two procedures give the same
result, and the result is consistent with the part enclosed by the first brace in (\ref{2ndAPP}).

To formally get rid of the second order time derivatives, first we can
get the Hamiltonian equation of motion with canonical formalism.
From Hamiltonian $H_\Phi=\int{d^3\vec{x}}\mathcal{H}$, where
$\mathcal{H}$ is given by (\ref{CompHamD}), we get
\bea\label{HamEqn}&&
\dot{\Phi}=\frac{\de{H_\Phi}}{\de\pi_\Phi}=-\frac{\pi_{\Phi^\dagger}}{e\tilde{G}^{00}}
-\frac{\tilde{G}^{0j}}{\tilde{G}^{00}}D_j\Phi-i[\frac{{k^0}^*_\phi}{\tilde{G}^{00}}-qA_0]\Phi,\nn&&
\dot{\pi}_{\Phi^\dagger}=-\frac{\de{H_\Phi}}{\de\Phi^\dagger}=
-D_j\left[\frac{\tilde{G}^{j0}}{\tilde{G}^{00}}\pi_{\Phi^\dagger}\right]
-i[\frac{{k^0}_\phi}{\tilde{G}^{00}}-qA_0]\pi_{\Phi^\dagger}\nn&&~~~
+D_i\left[e\bar{G}^{ij}D_j\Phi\right]
-e\left[m^2+\xi{R}+\hf{k_{\phi{A}}}^{\mn}F_{\mn}\right.\nn&&~~~
\left.-\frac{|k_\phi^0|^2}{\tilde{G}^{00}}\right]\Phi
+ie\left\{\left[k^j_\phi{D}_j\Phi+\frac{1}{e}D_j(e{k^j_\phi}^*\Phi)\right]\right.\nn&&~~~
\left.-\left[{k^0}_\phi\frac{\tilde{G}^{0j}}{\tilde{G}^{00}}D_j\Phi
+\frac{1}{e}D_j(e\frac{\tilde{G}^{j0}}{\tilde{G}^{00}}{k^0}^*_\phi\Phi)\right]
\right\}.
\eea
Then we can define $\Theta=+\frac{i}{m}\pi_{\Phi^\dagger}$ and symmetrize fields $\Phi,~\Theta$ by the definition
%\bea\label{Sym2CF}
$\Psi\equiv\left(\begin{array}{c}
            \eta \\
            \zeta
          \end{array}\right)\equiv
          \frac{1}{\sqrt{2}}\left(\begin{array}{c}
                              \Phi+\Theta \\
                              \Phi-\Theta
                            \end{array}\right).$
\begin{widetext}
For notational convenience, we can also define $\bar{g}^{\mn}\equiv{}g^{\mn}-K^{\mn}$,
and hence $\tilde{G}^{00}=\bar{g}^{00}$.
With these definitions, equation (\ref{HamEqn}) can be cast into the
Schr$\ddot{\mathrm{o}}$dinger form $i\dot{\Psi}=\hat{H}_\Psi\Psi$, where
\bea\label{HamLV02}&&
\hat{H}_\Psi
~~~=\left[\frac{a^0}{\bar{g}^{00}}-qA_0+\hf\nabla_j(\frac{S^{0j}}{\bar{g}^{00}})
+\frac{1}{2}\{\hat{\pi}_j,\frac{\bar{g}^{0j}}{\bar{g}^{00}}\}\right]\hat{1}
+\left[\hf\nabla_j(\frac{\bar{g}^{0j}}{\bar{g}^{00}})-\frac{b^0}{\bar{g}^{00}}-\frac{1}{2}\{\hat{\pi}_j,\frac{S^{0j}}{\bar{g}^{00}}\}\right]i\si_1
\nn&&~~~~~~~+\left\{\frac{e}{2m}\bar{M}^2+\frac{m}{2e\bar{g}^{00}}+\frac{1}{2m}\hat{\pi}_i(e\bar{G}^{ij}\hat{\pi}_j)-\hat{O}_k\right\}i\si_2
+\left\{\frac{e}{2m}M^2-\frac{m}{2e\bar{g}^{00}}+\frac{1}{2m}\hat{\pi}_i(e\bar{G}^{ij}\hat{\pi}_j)-\hat{O}_k\right\}\si_3.
\eea
In (\ref{HamLV02}), $\hat{\pi}\equiv[\hat{p}_i-qA_i]$,
$\hat{O}_k\equiv\frac{1}{2m}\left\{\left[\vec{\nabla}\cdot(e\vec{b})-\{\hat{\pi}_j,ea^j\}\right]
+\left[\nabla_j[e(a^0S^{0j}-b^0\bar{g}^{0j})/\bar{g}^{00}]+\{\hat{\pi}_j,\frac{e}{\bar{g}^{00}}(a^0\bar{g}^{0j}+b^0S^{0j})\}\right]
\right\}$, and $\si_i,~i=1,2,3$ are the Pauli matrices.
For completeness, up to now, we haven't assumed the isotropic metric and $A_\mu=0$, or done any approximation yet.
From the definition of pseudo-hermiticity
$\hat{\si}_3\hat{O}^\dagger\hat{\si}_3=\hat{O}$ \cite{FVWMP}, it it straightforward to verify that
$\hat{H}_\Psi$ in (\ref{HamLV02}) is pseudo-hermitian.
\end{widetext}
The pseudo-hermiticity requirement is necessary
to make sure all the eigenenergies of $\hat{H}_\Psi$ are real valued.
We also note the formal similarity of pseudo-hermiticity defined by $\si_3$ and that defined by $\ga^0$
in spinor space, \ie, $\ga^0\mathcal{M}^\dagger\ga^0=\mathcal{M}$.
This indicates that $\si_3$ plays a role very similar to $\ga^0$,
as can be seen from the prescription of dividing operators into even and odd parts in FWT \cite{FW}\cite{Case}.
In the following, we will take $A_\mu=0$ and the isotropic metric (\ref{SMetric}), so (\ref{HamLV02}) becomes
{\normalsize
\bea\label{RelHam1a}&&
\hat{H}_\Psi=
\left[\frac{a^0}{g^{00}}-\frac{K^{0j}}{g^{00}}\hat{p}_j
-\hf\hat{p}_j(\frac{\tilde{k}_{\phi\phi}^{j0}}{g^{00}})\right]\hat{1}
-\left[\frac{i}{2}\hat{p}_j(\frac{\tilde{k}_{\phi\phi}^{j0}}{g^{00}})\right.\nn&&~~
\left.+\frac{b^0}{g^{00}}-\frac{S^{j0}}{g^{00}}
\hat{p}_j\right]i\si_1
+\left\{\frac{e}{2m}M^2+\frac{m}{2e\bar{g}^{00}}\right.\nn&&~~
\left.+\frac{1}{2m}\hat{p}_i(e\bar{G}^{ij}\hat{p}_j)
-\hat{O}_k\right\}i\si_2
+\left\{\frac{e}{2m}M^2-\frac{m}{2e\bar{g}^{00}}\right.\nn&&~~
\left.+\frac{1}{2m}\hat{p}_i(e\bar{G}^{ij}\hat{p}_j)-\hat{O}_k\right\}\si_3,
\eea
}
where $M^2\equiv[m^2+\xi{R}]$,
$\hat{O}_k\equiv\frac{1}{2m}\left[\vec{\nabla}\cdot(e\vec{b})-\{\hat{p}_j,ea^j\}\right]$ and $\bar{g}^{00}=-(\frac{1}{V^2}+K^{00})$,
$\bar{G}^{ij}=\frac{\de_{ij}}{W^2}-\tilde{k}_{\phi\phi}^{ij},~e=VW^3$.
Note by replacing $\bar{g}^{00}$ with $g^{00}$ in the denominators, we have already ignored terms
with second order LV couplings.

\subsection{Pseudounitary transformation}
With the relativistic Hamiltonian (\ref{RelHam1a}), we can perform FWT directly to get the
NR approximation.
However, we wish to perform a pseudounitary transformation first, which will make the Hamiltonian more suitable for FWT,
then we do the FWT afterwards. We call this procedure the CVZ method, which was first introduced in \cite{CVZRW}.
For a similarity transformation to be defined as pseudounitary, its associated operator $\hat{U}$ must
satisfy $\hat{\si}_3\hat{U}^\dagger\hat{\si}_3=\hat{U}^{-1}$ \cite{CVZRW}\cite{Case}\cite{FVWMP}.
The goal of the desired pseudounitary transformation is to make the term proportional to $m$, $\frac{e}{2m}M^2+\frac{m}{2e\bar{g}^{00}}$, associated with $\si_2$, vanish.
Since the square brackets in (\ref{RelHam1a}) associated with $\hat{1}$ and $i\si_1$
do not contain any term proportional to $m$,
we can perform a ``rotation" only in the space spanned by $\si_2$ and $\si_3$, \ie,
define $\hat{U}\equiv{f+g\si_1}$ to eliminate the mass proportional term in the large brace multiplied by $i\si_2$.
Assuming $f,g\in\mathbb{R}^\infty$, the pseudounitary condition of $\hat{U}$ indicates $\hat{U}^{-1}=f-g\si_1$ and $f^2-g^2=1$.
With a little algebra, the mass-eliminating requirement gives
$\frac{f-g}{f+g}=e\sqrt{-\bar{g}^{00}}=W^3[1+\tilde{k}_{\phi\phi}^{00}V^2]^{\hf}$.
\begin{widetext}
Combined with $f^2-g^2=1$, we get
\bea\label{PUTO}&&
\hat{U}=
\hf(e\sqrt{-\bar{g}^{00}})^{-\hf}\left[1+e\sqrt{-\bar{g}^{00}}+(1-e\sqrt{-\bar{g}^{00}})\si_1\right].
\eea
Then we can use (\ref{PUTO}) to perform a pseudounitary transformation $\hat{H}'_\Psi\equiv\hat{U}^{-1}\hat{H}_\Psi\hat{U}$ on (\ref{RelHam1a}), which gives
\bea\label{PesuUni}&&
\hat{H}'_\Psi=
\left\{\frac{m}{\sqrt{-\bar{g}^{00}}}+\frac{\xi{R}}{2m\sqrt{-\bar{g}^{00}}}+
(e\sqrt{-\bar{g}^{00}})^{-\hf}[\frac{1}{2m}\hat{p}_i(e\bar{G}^{ij}\hat{p}_j)-\hat{O}_k](e\sqrt{-\bar{g}^{00}})^{-\hf}\right\}\si_3
\nn&&
~~~+\left\{\frac{\xi{R}}{2m\sqrt{-\bar{g}^{00}}}+
(e\sqrt{-\bar{g}^{00}})^{-\hf}[\frac{1}{2m}\hat{p}_i(e\bar{G}^{ij}\hat{p}_j)-\hat{O}_k](e\sqrt{-\bar{g}^{00}})^{-\hf}\right\}i\si_2
-\left[\frac{b^0}{g^{00}}+\frac{i}{2}\hat{p}_j(\frac{\tilde{k}_{\phi\phi}^{j0}}{g^{00}})+\frac{S^{0j}}{g^{00}}\hat{p}_j\right.\nn&&
~~\left.
+\frac{3}{2}\frac{K^{0j}}{g^{00}}\nabla_j\ln{W}\right]i\si_1+
\left\{\frac{a^0}{g^{00}}-\frac{K^{0j}}{g^{00}}\hat{p}_j
-\hf\hat{p}_j(\frac{\tilde{k}_{\phi\phi}^{j0}}{g^{00}})+\frac{3}{2}\frac{S^{0j}}{g^{00}}\nabla_j\ln{W}\right\}\hat{1}.
\eea

Following the spirit of FWT \cite{FW}\cite{Case}, we can separate $\hat{H}'_\Psi$ into even
and odd parts according to whether they commutate or anticommutate with $\si_3$,
where $\si_3$ plays the role of $\ga^0$ in the fermion case, as mentioned before.
In other words, we can write $\hat{H}'_\Psi=m\si_3+\mathcal{E}+\mathcal{O}$, where
$[\mathcal{E},\si_3]=0$ and $\{\mathcal{O},\si_3\}=0$.
Ignoring the non-minimal coupling term $\xi{R}$ and those which are products of the derivatives of $\chi$
and LV coefficients, the even and odd operators are
\bea&&\label{EvenO}
\mathcal{E}=\left\{
(e\sqrt{-\bar{g}^{00}})^{-\hf}\left[\frac{1}{2m}\hat{p}_i(e\bar{G}^{ij}\hat{p}_j)-\hat{O}_k\right]
(e\sqrt{-\bar{g}^{00}})^{-\hf}
%\right.\nn&&~~\left.
+m(\frac{1}{\sqrt{-\bar{g}^{00}}}-1)\right\}\si_3
+\left\{\frac{a^0}{g^{00}}-\frac{K^{0j}}{g^{00}}\hat{p}_j\right\}\hat{1},\\
&&\label{OddO}
\mathcal{O}=\left\{(e\sqrt{-\bar{g}^{00}})^{-\hf}\left[\frac{1}{2m}\hat{p}_i(e\bar{G}^{ij}\hat{p}_j)-\hat{O}_k\right]
(e\sqrt{-\bar{g}^{00}})^{-\hf}\right\}i\si_2
-\left[\frac{b^0}{g^{00}}+\frac{S^{0j}}{g^{00}}\hat{p}_j\right]i\si_1,
\eea
\end{widetext}
Now, clearly, $\mathcal{E}$ is already diagonal and hence decouples the two-component field $\Psi$,
while $\mathcal{O}$ is off-diagonal and still needs to be diagonalized.
In order to make the off-diagonal part smaller and smaller, we can perform a further unitary transformation
\bea\label{FFWT}
\Psi'\rightarrow{}A^{-1}\Psi',\quad \hat{H}'_\Psi{\rightarrow}A^{-1}{\hat{H}'_\Psi}A-iA^{-1}(\prt_tA),
\eea
where $A=\exp[-\frac{1}{2m}\si_3\mathcal{O}]$ \cite{Case}.
For a static metric, this transformation leads to
\bea\label{FWSca}&&
\hat{\tilde{H}}_\Psi=e^{\frac{1}{2m}\si_3\mathcal{O}}{\hat{H}'_\Psi}e^{-\frac{1}{2m}\si_3\mathcal{O}}\nn&&
~~~
=\hat{H}'_\Psi+\frac{1}{2m}[\si_3\mathcal{O},\hat{H}'_\Psi]+\frac{1}{8m^2}[\si_3\mathcal{O},[\si_3\mathcal{O},\hat{H}'_\Psi]]\nn&&~~
+\frac{1}{3!(2m)^3}[\si_3\mathcal{O},[\si_3\mathcal{O},[\si_3\mathcal{O},\hat{H}'_\Psi]]]+...\nn&&
~~=\si_3m+\left\{\mathcal{E}+\frac{1}{2m}\si_3\mathcal{O}^2-\frac{1}{8m^2}[\mathcal{O},[\mathcal{O},\mathcal{E}]]+...\right\}
\nn&&~~
+\left\{\frac{1}{2m}\si_3[\mathcal{O},\mathcal{E}]-\frac{1}{3m^2}\mathcal{O}^3+...\right\}.
\eea
Note that compared to $m$, all terms in $\mathcal{O},\mathcal{E}$ are either proportional to various powers of
the metric perturbation $\chi$ and its derivatives, or powers of tiny LV coefficients, or some products between the
two, which are all small parameters
(as mentioned before, in a weak gravitation field, $\hat{\vec{p}}^2/2m\sim{m}\chi\ll{m}$ can also be regarded as small).
So products of $\mathcal{O},\mathcal{E}$ must be much smaller, which legitimizes the approximation procedure
of the expansion in (\ref{FWSca}) \cite{SGI}.
Substituting the Schwarzschild metric $V=(1+\hf\chi)(1-\hf\chi)^{-1}$ and $W=(1-\hf\chi)^2$
into (\ref{EvenO},\ref{OddO}), and preserving only terms up to $\mathcal{O}(0,2), \mathcal{O}(1,1)$,
from (\ref{FWSca}), we get
\bea\label{NonRH}&&
\hat{H}_{_{\mathrm{CVZ}}}=\left\{m+\left[m\chi(1+\frac{\chi}{2}+\frac{\chi^2}{4})
-m\frac{\tilde{k}_{\phi\phi}^{00}}{2}(1+3\chi)\right]
\right.\nn
&&~~\left.
+\left[(1+3\chi+5\chi^2)-\frac{\tilde{k}_{\phi\phi}^{00}}{2}(1+5\chi)\right]\frac{\hat{\vec{p}}^2}{2m}
-\frac{(\hat{\vec{p}}^2)^2}{8m^3}\right.\nn&&
~~\left.+\frac{(1+\chi)}{2m}[2\vec{a}\cdot\hat{\vec{p}}-\tilde{k}_{\phi\phi}^{ij}\hat{\vec{p}}_i\hat{\vec{p}}_j]
-\frac{3}{4m}[2(\vec{\nabla}\chi)^2+\Delta\chi]\right.\nn
&&~~\left.-\frac{i}{2m}(3+10\chi)\vec{\nabla}\chi\cdot\hat{\vec{p}}\right\}\si_3
%\nn&&~~
+(1+2\chi)[K^{0j}\hat{p}_j-a^0]\hat{1}.\nn
\eea
Note that $-\frac{(\hat{\vec{p}}^2)^2}{8m^3}$ comes from the lowest-order LI contribution of $\frac{1}{2m}\si_3\mathcal{O}^2$,
and all the other terms except $m$ come from $\mathcal{E}$. Up to $\mathcal{O}(1,1),\mathcal{O}(0,2)$,
we haven't even calculated $-\frac{1}{8m^2}[\mathcal{O},[\mathcal{O},\mathcal{E}]]$.
Compared to direct FWT which will be shown below,
we see that the pseudounitary transformation saves the work of calculating commutators in (\ref{FWSca}),
if the NR approximation is only required to proceed to next leading order.
As mentioned before, except the last two terms in the large brace, (\ref{NonRH}) agrees well with the terms in the
first brace of (\ref{2ndAPP}), indicating that it is still possible to improve the NR procedure using the conventional method.

\subsection{Foldy-Wouthuysen transformation}
In this subsection, we show that direct FWT on (\ref{RelHam1a}) can also
lead to the same result in (\ref{NonRH}).
For calculational convenience, we can separate both $\mathcal{E}$ and $\mathcal{O}$ into LI and LV parts,
\ie,
$\mathcal{E}=\mathcal{E}_{LI}+\mathcal{E}_{LV}$ and
$\mathcal{O}=\mathcal{O}_{LI}+\mathcal{O}_{LV}$.
In detail,
\bea&&\label{EvenLI}
\mathcal{E}_{LI}=
\left\{\frac{em}{2}-\frac{m}{2eg^{00}}-m+\frac{1}{2m}\hat{p}_i(VW\hat{p}_i)\right\}\si_3,
%\nn&&
%=\left\{m\chi(1+5\chi+7\chi^2)+\frac{1}{2m}[(1-\frac{\chi^2}{4})\hat{\vec{p}}^2
%+\frac{i\chi}{2}\vec{\nabla}\chi\cdot\hat{\vec{p}}]\right\}\si_3,\nn
\\&&\label{EvenLV}
\mathcal{E}_{LV}=\left\{-\frac{m}{2}\tilde{k}_{\phi\phi}^{00}\mathcal{F}^3+
\frac{VW^3}{2m}\left[2\vec{a}\cdot\hat{\vec{p}}-\tilde{k}_{\phi\phi}^{ij}\hat{p}_i\hat{p}_j\right]\right\}\si_3
\nn&&~~~~~+\left\{V^2[K^{0j}\hat{p}_j-a^0]\right\}\hat{1},
%\nn&&~~
%=\left\{-\frac{m\tilde{k}_{\phi\phi}^{00}}{2}(1+6\chi)+
%(1-2\chi)[\frac{\vec{a}\cdot\hat{\vec{p}}}{m}-\frac{\tilde{k}_{\phi\phi}^{ij}}{2m}\hat{p}_i\hat{p}_j]\right\}\si_3
%\nn&&~~+\left\{(1+2\chi)[K^{0j}\hat{p}_j-a^0]\right\}\hat{1},
\eea
%and
\bea&&\label{OddLI}
\mathcal{O}_{LI}=
\left\{\frac{em}{2}+\frac{m}{2eg^{00}}+\frac{1}{2m}\hat{p}_i(VW\hat{p}_i)\right\}i\si_2,
%\nn&&
%=\left\{-3m\chi(1+\frac{5}{4}\chi+\frac{7}{3}\chi^2)+\frac{1}{2m}[(1-\frac{\chi^2}{4})\hat{\vec{p}}^2
%+\frac{i\chi}{2}\vec{\nabla}\chi\cdot\hat{\vec{p}}]\right\}i\si_2,\nn
\\&&\label{OddLV}
\mathcal{O}_{LV}=\left\{\frac{m}{2}\tilde{k}_{\phi\phi}^{00}\mathcal{F}^3+
\frac{VW^3}{2m}\left[2\vec{a}\cdot\hat{\vec{p}}-\tilde{k}_{\phi\phi}^{ij}\hat{p}_i\hat{p}_j\right]\right\}i\si_2
\nn&&~~~~~+\left\{V^2[b^0+S^{0j}\hat{p}_j]\right\}i\si_1.
%\nn&&~~
%=\left\{\frac{m\tilde{k}_{\phi\phi}^{00}}{2}(1+6\chi)+
%(1-2\chi)[\frac{\vec{a}\cdot\hat{\vec{p}}}{m}-\frac{\tilde{k}_{\phi\phi}^{ij}}{2m}\hat{p}_i\hat{p}_j]\right\}i\si_2
%\nn&&~~+\left\{(1+2\chi)[b^0+S^{0j}\hat{p}_j]\right\}i\si_1.
\eea
So expanded in terms of $\chi$ and its derivatives,
we have up to linear order of LV coefficients,
\bea\label{SquO}&&
\mathcal{O}^2=\mathcal{O}_{LI}^2+\{\mathcal{O}_{LI},\mathcal{O}_{LV}\}
=-\hat{1}\left\{9m^2\chi^2(1+\frac{5}{2}\chi)\right.\nn&&~~
\left.+(\frac{\hat{\vec{p}}^2}{2m})^2-3\chi(1+\frac{5}{4}\chi)\hat{\vec{p}}^2
+3i(1+\frac{5}{2}\chi)\vec{\nabla}\chi\cdot\hat{\vec{p}}
\right.\nn&&~~
\left.+\frac{3}{2}[\Delta\chi+\frac{5}{2}(\vec{\nabla}\chi)^2]
+\frac{\tilde{k}_{\phi\phi}^{00}}{2}(1+6\chi)\hat{\vec{p}}^2-3m^2\tilde{k}_{\phi\phi}^{00}\chi
\right.\nn&&~~
\left.+\left[(1-2\chi)\frac{\hat{\vec{p}}^2}{m^2}-6\chi\right]
\left[\vec{a}\cdot\hat{\vec{p}}-\frac{\tilde{k}_{\phi\phi}^{ij}}{2}\hat{p}_i\hat{p}_j\right]\right\},
\eea
and
\bea\label{OEOC}&&
\left[\mathcal{O},[\mathcal{O},\mathcal{E}]\right]=\left[\mathcal{O}_{LI},[\mathcal{O}_{LI},\mathcal{E}_{LI}]\right]
+\left[\mathcal{O}_{LI},[\mathcal{O}_{LI},\mathcal{E}_{LV}]\right]\nn&&~~
+\left[\mathcal{O}_{LI},[\mathcal{O}_{LV},\mathcal{E}_{LI}]\right]
+\left[\mathcal{O}_{LV},[\mathcal{O}_{LI},\mathcal{E}_{LI}]\right],
\eea
where
\bea\label{3rdLI}&&
\left[\mathcal{O}_{LI},[\mathcal{O}_{LI},\mathcal{E}_{LI}]\right]=\nn&&~~~
\left\{
\left[(5\chi\hat{\vec{p}}^2-8i\vec{\nabla}\chi\cdot\hat{\vec{p}}
-\Delta\chi)\right]\frac{\hat{\vec{p}}^2}{m}-\hf(\frac{\hat{\vec{p}}^2}{m})^3\right.\nn&&~~~
\left.+6m\chi\left[2(i\vec{\nabla}\chi\cdot\hat{\vec{p}}+\Delta\chi)-6m^2\chi^2-\chi\hat{\vec{p}}^2\right]
\right\}\si_3,\nn
\eea
\bea\label{3rdLV}&&
\left[\mathcal{O}_{LI},[\mathcal{O}_{LI},\mathcal{E}_{LV}]\right]
+\left[\mathcal{O}_{LI},[\mathcal{O}_{LV},\mathcal{E}_{LI}]\right]\nn&&~~~
+\left[\mathcal{O}_{LV},[\mathcal{O}_{LI},\mathcal{E}_{LI}]\right]=
\left\{\frac{\tilde{k}_{\phi\phi}^{00}}{4}\chi\frac{\hat{\vec{p}}^2}{m}
+\frac{\tilde{k}_{\phi\phi}^{00}}{16}(\frac{\hat{\vec{p}}^2}{m})^2\right.\nn&&~~~
\left.
-\frac{17\chi}{8m}[\frac{\vec{a}\cdot\hat{\vec{p}}}{m}
-\frac{\tilde{k}_{\phi\phi}^{ij}}{2m}\hat{p}_i\hat{p}_j]\frac{\hat{\vec{p}}^2}{m}\right\}\si_3.
\eea
Substituting all the above equations (\ref{EvenLI}-\ref{3rdLV}) back into
(\ref{FWSca}), we get the NR scalar Hamiltonian (up to second order commutators of FWT) as
\bea\label{NonRHDFW}&&
\hat{H}_{_\mathrm{FWT}}=\left\{m+m\chi(1+\frac{\chi}{2}+\frac{\chi^2}{4})-\frac{i}{2m}(3+10\chi)\vec{\nabla}\chi\cdot\hat{\vec{p}}\right.\nn&&~~
\left.+\left[(1+3\chi+5\chi^2)-\frac{\tilde{k}_{\phi\phi}^{00}}{2}(1+5\chi)\right]\frac{\hat{\vec{p}}^2}{2m}-\frac{(\hat{\vec{p}}^2)^2}{8m^3}
\right.\nn&&~~
\left.-m\frac{\tilde{k}_{\phi\phi}^{00}}{2}(1+3\chi)
+\frac{(1+\chi)}{2m}[2\vec{a}\cdot\hat{\vec{p}}-\tilde{k}_{\phi\phi}^{ij}\hat{\vec{p}}_i\hat{\vec{p}}_j]\right.\nn&&~~
\left.-\frac{3}{4m}\Delta\chi-(1+\frac{9}{4}\chi)[2\vec{a}\cdot\hat{\vec{p}}-\tilde{k}_{\phi\phi}^{ij}\hat{\vec{p}}_i\hat{\vec{p}}_j]
\frac{\hat{\vec{p}}^2}{4m^3}\right\}\si_3
\nn&&~~
+(1+2\chi)[K^{0j}\hat{p}_j-a^0]\hat{1}.
\eea
Compared with (\ref{NonRH}), we see that except for the LV term proportional to $\frac{\hat{\vec{p}}^2}{4m^3}$,
the NR Hamiltonian obtained by direct FWT is completely
the same as that obtained with CVZ method, though to the next lowest order,
the latter can be obtained without substantially calculating any commutators.
At first glance, this is a little surprising, because the results are expected to differ by a pseudounitary transformation,
however, inspecting the CVZ method, we see that it is exactly the pseudounitary transformation
which ensures the NR Hamiltonian is the same as that obtained with direct FWT \cite{SilenkoC}.
Since the pseudounitary transformation preserves both the charge and matrix elements of the Hamiltonian after transformation \cite{FVWMP}.

\subsection{Consistency check and partial support}
Another confirmation can be seen by applying the different methods mentioned above to the linear accelerating metric
$g_{00}=-[1+\frac{\vec{a}\cdot\vec{x}}{c^2}]^2,~g_{ij}=\de_{ij}$. With either direct FWT, CVZ method,
or even the unsystematically traditional method in Sec.\ref{TNRKG}, we can get an NR Hamiltonian
\bea\label{NRHLA}&&
\hat{H}_{_\mathrm{NRL}}=m(1+\phi)+\left[(1+\phi)-\frac{\tilde{k}_{\phi\phi}^{00}}{2}(1+3\phi)\right]\frac{\hat{\vec{p}}^2}{2m}\nn&&~~
-\frac{\tilde{k}_{\phi\phi}^{00}}{2}m(1+3\phi)
+(1+\phi)\left[\frac{\vec{a}\cdot\hat{\vec{p}}}{m}-\frac{\tilde{k}_{\phi\phi}^{ij}}{2m}\hat{\vec{p}}_i\hat{\vec{p}}_j\right]
\nn&&~~
-\frac{i}{2m}\vec{\nabla}\phi\cdot\hat{\vec{p}}
+(1+2\phi)[K^{0j}\hat{p}_j-a^0]-\frac{(\hat{\vec{p}}^2)^2}{8m^3},\nn
\eea
where $\phi\equiv\frac{\vec{a}\cdot\vec{x}}{c^2}$. The correctness of (\ref{NonRHDFW}) and (\ref{NRHLA}) can
be partially supported by comparing the LI part of these Hamiltonians with the equations (20) and (21) in \cite{ABFW}.
We can even compare the LI part of (\ref{NRHLA}) with the fermion Hamiltonian obtained in \cite{YuriEPI}\cite{SGI}\cite{WTNH}.
The consistency of this comparison lies in the fact that each spinor component
satisfies the Klein-Gordon equation as dictated by the relativistic dispersion relation.
In other words, the NR Hamiltonian for a scalar field is equivalent to that of a fermion field by ignoring
its spin contribution.
In the same spirit, we can also expect an equivalence between the LV contribution to the NR scalar Hamiltonian with the fermion counterpart,
see \cite{MattGC}\cite{YuriEPI}.
Comparing (\ref{NRHLA}) with (21,23,27) in \cite{YuriEPI} by ignoring the spin interactions, we find
\bea\label{LVC}&&
(k_\phi)^0\sim[\tilde{a}^0-me^0],\quad (k_\phi)^j\sim[\tilde{a}^j-me^j],\nn&&
\frac{\tilde{k}_{\phi\phi}^{00}}{2}\sim{c}_{00},\quad
\frac{\tilde{k}_{\phi\phi}^{ij}}{2}\sim{c}_{(ij)},\quad K^{0i}\sim2{c}_{(0i)}.
\eea
Note that to avoid notational confusion of the real part of $(k_\phi)^\mu$, \ie, $a^\mu$, with the fermionic LV
coefficient ``$a^\mu$", we instead use $\tilde{a}^\mu$ to represent the latter in (\ref{LVC}).
The relation (\ref{LVC}) is also consistent with the CPT properties of the corresponding LV coefficients.
The incompleteness of the formal similarity between the spin-independent NR LV Hamiltonian (\ref{NRHLA})
and the one in \cite{YuriEPI} can be attributed to the fact that we only preserve LV perturbations to $\mathcal{O}(1,1)$
for simplicity,
while in \cite{YuriEPI} these perturbations are preserved to much higher orders.
An interesting scenario is that if we start
with a Lagrangian describing spin-1 boson (such as meson) and carry out the above procedure again, we may also establish
a relationship between the spin-dependent LV coefficients for an effective boson with
the more fundamental fermionic LV couplings, like $b^\mu,~H^{\mn},~g^{\la\mn}$, \etc

Finally, we mention again the advantage of the CVZ method over direct FWT is that,
at least to the next lowest order of the NR approximation,
the CVZ method can largely save work in calculating various commutators, such as $\left[\mathcal{O},[\mathcal{O},\mathcal{E}]\right]$ in FWT.

\section{Relation to the Test of the Equivalence Principle}\label{EPSNR}
Next, we'd like to discuss the relevance of the scalar Hamiltonian to the test of the
equivalence principle (EP). Actually, there are various inequivalent definitions on EP
in the intensive discussions found in the literature \cite{EPIS}.
Thus it is no doubt that discussions of inequivalent subjects
necessarily cause conflicting conclusions on the validity of EP \cite{Ahluwalia}\cite{OkonEP}.
As mentioned at the very beginning, we constrain ourselves to the WEP. Speaking more precisely,
we mean the equivalence between the law of mechanics for any free-moving test body with negligible self-gravity in a
sufficiently small local region of spacetime (in a gravitational field) with that in a uniform accelerating frame
(with proper acceleration) in the absence of gravity \cite{EPIS}\cite{PoiWill}.
Note that in this statement, universal free fall (UFF, the world line of any free-moving test body with given initial conditions
is independent of its mass and internal properties.) cannot be equivalent to WEP \cite{QEPMS},
and ceases to be valid in quantum domain. More seriously, UFF is even meaningless in
quantum mechanics as the world line (or trajectory for an object) is purely a classical concept.
In this sense, it is better to view UFF as a classical manifestation of WEP. On the contrary, WEP can still be
safely guaranteed in quantum realm, especially constrained to NR region reduced \cite{OkonEP}\cite{EPOK}\cite{SPWEP}
from GR. Actually, WEP provides a key to ``gauge away" the gravitational analogy of gauge potential, the first derivatives of metric tensor, \ie, $\prt_\rho{g}_{\mn}\sim\Ga_{\rho\mn}$, thus is an essential ingredient to glue quantum matter (neglecting spin-gravity couplings) to the classical gravitational background.
In relativistic quantum field theory, WEP may not be valid due to the non-local nature of the radiative corrections even in a classical GR background \cite{QEDQG}.

In this respect, we think it is more meaningful to test WEP in an extended theory of GR,
especially in the quantum domain. Many alternative theories fit into this category,
like Einstein-Cartan theory \cite{Cartan}, metric-affine theory \cite{MAG}, etc.
In a much broad context, it is valuable to incorporate Lorentz and CPT violation
together with the test of WEP in a single framework, especially considering
the intimate relationship between LLI and WEP, as indicated by Schiff's conjecture \cite{EPCW}\cite{Schiff}.
SME provides such an ideal test ground.
In fact, testing WEP in SME allows more exotic signals, like the distinctive nature between
gravitational force and acceleration in the presence of LV \cite{MattGC}.
Discussion of EP in the context of SME is abundant \cite{YuriEP}\cite{YuriEPI}\cite{EXPEPL}\cite{MattGC}\cite{QBailey},
however, it seems that two important points have been overlooked or not been taken seriously,
which we'd like to stress below.

First, it is logically more consistent to start with an intrinsically curved metric
instead of a uniform accelerating metric, though % where tidal effects can be neglected,
the latter is an excellent approximation in most circumstances (up to an irrelevant constant),
\eg, $g_{00}\simeq-(1+2\chi)=-[1+2\vec{g}\cdot\Delta\vec{r}/c^2+2\frac{GM}{c^2R}]%\sim-(1+\vec{g}\cdot\Delta\vec{r}/c^2)^2
\sim-(1+2\phi)$ ($R$ is the Earth radius).
However, this approximation cannot be reliable to higher orders.
In essence, the metric $g_{00}=-(1+\vec{a}\cdot\vec{x}/c^2)^2,~g_{ij}=\de_{ij}$ is only
a general relativistic description of uniform acceleration, which is essentially flat, and contains no information of gravity.
Comparing (\ref{NonRHDFW}) with (\ref{NRHLA}), we see, even staying at the metric level and in the absence of LV,
the two Hamiltonians cannot be equivalent at orders other than $\mathcal{O}(\chi)$,
not mention the $\vec{\nabla}\chi\cdot\hat{\vec{p}}$ or $\vec{\nabla}\phi\cdot\hat{\vec{p}}$ term.
Viewing in another way, the failure of this match may precisely reflect the realm of validity in the statement of WEP,
``a sufficiently small local region of spacetime". Going beyond this ``local" patch of spacetime necessarily means
going out of the domain of WEP, where ``violation" is naturally expected even in GR.

%{\cred (here Ralf's suggestion have to be taken into account!)}
Second, speaking about how small should be considered as local enough, that depends on experimental capabilities.
For an experimental apparatus capable of achieving the precision of $\de{s}~\mu\text{gal}$ in
a gravitational acceleration measurement near the Earth surface, the length scale is roughly about
$L=L(\de{s})\sim\frac{\de{|\vec{g}|}_\mathrm{max}R}{2g}\sim10^{-8}\times\frac{\de{s}R}{2g}$ to
ensure the local requirement of WEP test, otherwise even the conventional tidal gravity can have a nonnull effect.
For example, if the gravimeter precision is of order $1\text{mgal}$, the length scale involved in
the gravimeter measurement must be less than $3.24$ m, which is easy to satisfy.
For a $1\mu\text{gal}$ precision measurement, the length scale is smaller by a factor of a thousand,
which excludes many conventional macroscopic gravitational experiments.

On the other hand, if the flavor-dependent LV coefficients $\tilde{k}_{\phi\phi}^{\mn},~\tilde{k}_\phi^\mu$
are nonzero, WEP is apparently violated.
To see this, we collect the LV Hamiltonian up to $\mathcal{O}(\chi)$ from (\ref{NonRHDFW}) as below,
\bea\label{LNonRH}&&
\hat{H}_{_\mathrm{FWT}}=\left[m\chi(1-\frac{3\tilde{k}_{\phi\phi}^{00}}{2}-2\frac{a^0}{m})
-\frac{m\tilde{k}_{\phi\phi}^{00}}{2}-a^0\right]
\nn&&~~
+\left(1-\frac{\tilde{k}_{\phi\phi}^{00}}{2}\right)\frac{\hat{\vec{p}}^2}{2m}
+(1+2\chi)K^{0j}\hat{p}_j
\nn&&~~
+\frac{(1+\chi)}{2m}\left[2\vec{a}\cdot\hat{\vec{p}}-\tilde{k}_{\phi\phi}^{ij}\hat{\vec{p}}_i\hat{\vec{p}}_j\right].
\eea
%$\left[m\chi(1-\frac{3\tilde{k}_{\phi\phi}^{00}}{2})-\frac{m\tilde{k}_{\phi\phi}^{00}}{2}-(1+2\chi)a^0\right]$
The first term in the large square bracket can be regarded as potential energy depends
not only on the LV corrected mass term $m(1-\frac{3\tilde{k}_{\phi\phi}^{00}}{2}-2\frac{a^0}{m})$,
but also directly on the combination of LV coefficients, $-[m\tilde{k}_{\phi\phi}^{00}/2+a^0]$.
In general, the LV coefficients are directionally dependent, and hence
necessarily lead to breaking of UFF even in the context of classical mechanics.
We can see this more transparently from the classical Lagrangian (\ref{ClaLa}) derived below.
In fact, even performing the usual coordinate transformation
$z\rightarrow{z'=z+\frac{g}{2}t^2},~t\rightarrow{t'=t}$ on the Schr$\ddot{\mathrm{o}}$dinger
equation \cite{EPOK} associated with Hamiltonian (\ref{LNonRH}), it cannot be reduced to
the free motion case even locally ($\chi\rightarrow\vec{g}\cdot\Delta\vec{r}/c^2$)
due to the presence of LV coefficients.
So LV necessarily violates WEP by definition.
Inspection of (\ref{NonRHDFW}) also reveals that gravitational redshift
associated with $\tilde{k}_{\phi\phi}^{\mn}$ depends on the number of its zero indices, so
this can be utilized to discriminate different LV coefficients, as already been noticed
in \cite{YuriEPI}. This also prevents us from using a coordinate transformation to the local
patch of uniform acceleration frame, to transform Hamiltonian (\ref{LNonRH}) to
the flat space one with LV couplings. %(we already ignored various fluctuations of LV
%coefficients in the above discussions).

To see violation of WEP in another way, from the quadratic dispersion relation
\bea\label{LDSP}&&
(\frac{1}{V^2}+\tilde{k}_{\phi\phi}^{00}){p_0}^2+(\tilde{k}_{\phi\phi}^{ij}-\frac{\de^{ij}}{W^2})p_ip_j
+\tilde{k}_{\phi\phi}^{(0i)}p_ip_0\nn&&~~
+\frac{i}{W^2}\nabla_i\ln(VW)p_i-2(a^0p_0+a^jp_j)=m^2\nn
\eea
derived from (\ref{SEOM}), we can construct a classical relativistic Lagrangian \cite{ARCL}
\bea\label{ClaLa}&&
L=-\mu\left[V^2(1-\tilde{k}_{\phi\phi}^{00}V^2){u^0}^2-W^2(\de_{ij}+K_{ij}W^2)u^iu^j\right.
\nn&&~~
\left.+2V^2W^2K_{0i}u^0u^i\right]^{\hf}+\left[W^2a_j-\frac{i}{2}\nabla_j\ln(VW)\right]u^j\nn&&~~
-a_0V^2u^0,
\eea
where $\mu\equiv\left\{m^2+\frac{1}{4W^2}[\vec{\nabla}\ln(VW)]^2\right\}$, $K^{\mn}\equiv{\mathrm{Re}}[\tilde{k}_{\phi\phi}^{\mn}]=K^{\nu\mu},~u^\mu\equiv\frac{dx^\mu}{d\tau}$.
As a simple approximation, we have only retained the LV coefficients in the above calculation to linear order.
It can be readily verified that the particle trajectory obtained from (\ref{ClaLa}) deviates from
geodesic equation $\frac{du^0}{d\tau}+2(\vec{u}\cdot\vec{\nabla}\ln{V})u^0=0$,
$\frac{d\vec{u}}{d\tau}+\hf\frac{\vec{\nabla}{V^2}}{W^2}{u^0}^2+2(\vec{u}\cdot\vec{\nabla}\ln{W})\vec{u}
-\vec{u}^2\vec{\nabla}\ln{W}=0$ without LV, and hence apparently violates WEP classically, \ie, UFF.
Note that (\ref{ClaLa}) is only a toy illustration to show that the inclusion of LV necessarily indicates
deviation from geodesic for a classical particle trajectory. Since the equation of motion derived from (\ref{ClaLa})
automatically includes various products of LV coefficients with $\prt_ig_{00}$ or $\prt_ig_{jk}$, to be
self-consistent, we have to include higher order LV contributions as well, which is beyond the scope of
this paper.

At the end of this section, we note that there are several subtleties in the discussion of WEP.
One issue is that, the non-local nature of vacuum polarization may induce non-minimal couplings even
starting with a minimal coupled action \cite{QEDQG}, as mentioned before.
This effect can introduce a very tiny length scale, the Compton wavelength $\la_C$ of a massive particle, say, the electron,
and this will definitely violate WEP due to the tidal effects.
The other issue is particular for the presence of LV, the so-called vacuum Cherenkov radiation \cite{VCRP}\cite{VCRCS}\cite{VCRF}\cite{VCRWZG}.
For an energetic charged particle whose velocity exceeds the phase velocity of LV photon, the charge is expected to radiate \cite{VCRP}\cite{VCRCS}.
Similarly, a Cherenkov-type process can occur for modified electroweak and gravity sectors as well, leading to the emission
of W, Z bosons and gravitons, respectively \cite{VCRWZG}.
The back-reaction due to this radiation can lead to a deviation from geodesic motion \cite{VCRP}, however, except for the electromagnetic Maxwell-Chern-Simons
theory \cite{VCRCS}, due to the existence of threshold energy, this scenario will be non-relevant for a NR particle in general.
While for the LV charged fermion, the situation is a little complicated.
Certain spin-flip LV coefficients like $H$, $d$ and $g$ can also lead to threshold-free vacuum Cherenkov radiation \cite{VCRF},
and this will drive even a NR charged particle away
from its geodesic. For an effective neutral particle composed of charged fermions, it is still unclear whether the composite charged fermions
in bound state can radiate or not. If they can, the back-reaction may lead to WEP violation as well, though this could be a higher-order LV effect.

\section{Summary}\label{Summary}
In this work, we have derived a NR gravitationally coupled scalar Hamiltonian from the scalar
Lagrangian of minimal SME. Using the test particle assumption, we derive it from two different methods in a
static isotropic metric. One derivation utilizes the usual ansatz $\Phi(t,\vec{r})=e^{-imt}\Psi(t,\vec{r})$.
The other is the Foldy-Wouthuysen transformation (FWT) with a pseudounitary transformation developed by Cognola \etl~\cite{CVZRW},
and we call it the CVZ method.
At least to $\mathcal{O}(1,1)$, the results (\ref{2ndAPP}) and (\ref{NonRH}), obtained from the two different
methods match.
In the former method, we used iteration procedure to perturbatively eliminate additional time derivative terms
like $\frac{\ddot{\psi}}{2m}$, which proves to be crucial for correct approximation.
This method is a bit loose, though we think it is much more straightforward, and
it will be interesting to explore whether this method can be further developed to obtain higher-order corrections systematically.
We also check the CVZ method with a direct FWT, and the result (\ref{NonRHDFW}) confirms (\ref{NonRH}) very well.
However, at least for the next-leading-order approximation, the CVZ method appears more economical, as it largely saves
the work in calculating various commutators.

In the context of SME, various NR Hamiltonians stemming from fermion Lagrangian have been developed in the literature \cite{LVNRH}\cite{Nonmini}.
It is natural because matter is composed of fermions.
However, in an effective point of view, it is complementary to start directly with a bosonic action,
since many quantum tests of WEP began to use bosonic atoms \cite{EPTheo}\cite{bosonA0}\cite{bosonA}
as test particles.
Our result provides such an example for the spin-0 boson, which may be useful to the analysis of the $^{88}$Sr atom \cite{bosonA0}.
Generalization to the spin-1 case will be straightforward, and may be more interesting since spin interaction
allows experimental testing in a more general framework, like metric-affine theory
with torsion and nonmetricity \cite{ExtGLI}\cite{ExtGLV},
so more broad test schemes \cite{SpinWPT}\cite{SPINTEST} are involved. As a bonus, comparison of
NR Hamiltonian for scalar and fermion fields enables us to bridge a relation between the corresponding
LV coefficients, see (\ref{LVC}). Accordingly, we may also be able to establish a relation between
the LV coefficients of the spin-1 boson field and those of the fermion field in a future work. Then the spin-dependent
LV coefficients, like $H^{\mn},~d^{\mn},~g^{\la\mn}$ may be able to match the counterparts of spin-1
boson, which is not attainable in the scalar case.

Finally, we also discuss the relevance of the scalar Hamiltonian with the test of WEP,
which in our conservative point of view is still valid in the semi-classical context
in the nonrelativistic regime reduced from GR.
So tests of WEP are much natural in an extended theory of GR.
With both a classical Lagrangian and a NR Hamiltonian, we show that classically, the presence of LV
indeed leads to deviation of the geodesic, which is apparently a signal of UFF violation.
Furthermore, since the LV coefficients are directionally dependent, and receive gravitational redshift differently,
we argue that this also leads to breaking of WEP even when transformed to a uniform accelerating frame with $\vec{a}=-\vec{g}$.
Speically, If LV leads to vacuum Cherenkov radiation, due to the back-reaction of the emitted quanta to test particle,
more subtle WEP violation effects are expected for a composite neutral scalar.

\section{Acknowledgement}
The author is gratiful to Chao-guang Huang, Ralf Lehnert, Marco Schreck for valuable discussions, and Alan Kosteleck\'y for helpful suggestions,
and also would like thanks the anonymous referee for very helpful comments, pointing out that the back-reaction of vacuum Cherenkov radiation
can also lead to WEP violation.
This work is also partially supported by the National Natural Science Foundation of China under grant No. 11605056, No. 11574082,
and the Fundamental Research Funds for the Central Universities under No. 2017MS052.


\begin{thebibliography}{99}
\bibitem{HiggsLHC}The CMS Collaboration, \Journal{\NP}{10}{2014}{557}.
\bibitem{GravWave}B.P. Abbott {\it et al.}, \Journal{\PRL}{116}{2016}{061102};
ibid, \Journal{\PRL}{116}{2016}{241103}£»ibid, \Journal{\PRL}{118}{2017}{221101}£»
ibid, \Journal{\PRL}{119}{2017}{161101}.
\bibitem{FireWall}A. Almheiri, D. Marolf, J. Polchinski and James Sully, \Journal{\JHEP}{02}{2013}{062};
P. Chen, Y.C. Ong, D.N. Page, M. Sasaki and Dong-han Yeom, \Journal{\PRL}{116}{2016}{161304}.
\bibitem{Update}S. Liberati, \Journal{\CQG}{30}{2013}{133001}; \Journal{\JPCS}{631}{2015}{012011}; J.D. Tasson, \Journal{\RPP}{77}{2014}{062901}.
\bibitem{DATA}V.A. Kosteleck\'y and N. Russell, \Journal{\RMP}{83}{2011}{11}.
\bibitem{PLV}V. A. Kosteleck\'y and S. Samuel, \Journal{\PRL}{63}{1989}{224}; ibid. \Journal{\PRL}{66}{1991}{1811}; ibid. \Journal{\PRD}{39}{1989}{683};
    R. Gambini and J. Pullin,  \Journal{\PRD}{59}{1999}{124021};
    J. Alfaro, H. A. Morales-Tecotl, and L. F. Urrutia, \Journal{\PRL}{84}{2000}{2318};
    G. Amelino-Camelia, J. R. Ellis, N. E. Mavromatos, D. V.
Nanopoulos, and S. Sarkar, \Journal{\NT}{393}{1998}{763}.
\bibitem{sme}D. Colladay and V.A. Kosteleck\'y, \Journal{\PRD}{55}{1997}{6760}; \Journal{\PRD}{58}{1998}{116002}.
\bibitem{2004Alan}V.A. Kosteleck\'y, \Journal{\PRD}{69}{2004}{105009}.
\bibitem{Nonmini}V.A. Kosteleck\'y and M. Mewes, \Journal{\PRD}{80}{2009}{015020}; \Journal{\PRD}{85}{2012}{096005}; \Journal{\PRD}{88}{2013}{096006}.
\bibitem{CPTM}V.A. Kosteleck\'y (editor), CPT and Lorentz Symmetry, Proceedings of the Fifth-Seventh Meeting on CPT and Lorentz Symmetry, (World Scientific, Singapore, 2010, 2011, 2016).
\bibitem{EPTheo}T. Damour, \Journal{\CQG}{13}{1996}{A33}; M. Zych and C. Brukner, arXiv:1502.00971[gr-qc];
G. Rosi, G.D¡¯Amico, L. Cacciapuoti, F. Sorrentino, M. Prevedelli, M. Zych, C¡¦. Brukner
and G.M. Tino, \Journal{\NC}{8}{2017}{15529}.
\bibitem{EPExp}V.V. Flambaum, \Journal{\PRL}{117}{2016}{072501}.
\bibitem{TorsBan2012}T.A. Wagner, S. Schlamminger, J.H. Gundlach and E.G. Adelberger,
\Journal{\CQG}{29}{2012}{184002}.
\bibitem{Shao2016}L. Shao and N. Wex, \Journal{\SCPMA}{59}{2016}{699501}.
\bibitem{EPCW}C.M. Will, \Journal{\LRR}{17}{2014}{4}.
\bibitem{SWLSEP}S. Weinberg, \Journal{\PL}{9}{1964}{357}; The Quantum Theory of Fields, Cambridge University Press, Chap. 13.1, Page 537.
\bibitem{MapEPLV}A. Halprin and H.B. Kim, \Journal{\PLB}{469}{1999}{78}.
\bibitem{Schiff}L. Schiff, \Journal{\AJP}{28}{1960}{340}.
\bibitem{Ahluwalia}G.Z. Adunas, E. Rodriguez-Milla and D. V. Ahluwalia, \Journal{\GRG}{33}{2001}{183};
P.J. Orlando, R.B. Mann, K. Modi and F.A. Pollock, \Journal{\CQG}{33}{2016}{19LT01}.
\bibitem{OkonEP}E. Okon and C. Callender, \Journal{\EJPS}{1}{2011}{133}.
\bibitem{EPOK}M. Nauenberg, \Journal{\AJP}{84(11)}{2016}{979}; H. Padmanabhan and T. Padmanabhan, \Journal{\PRD}{84}{2011}{085018}.
\bibitem{Barg}V. Bargmann, \Journal{\AoM}{59}{1954}{1-46};
S. Weinberg, The Quantum Theory of Fields, Cambridge University Press, Chap. 2.4 Page 62.
\bibitem{MicroscopeM} P. Touboul {\it et al.}, \Journal{\PRL}{119}{2017}{231101}.
\bibitem{SpinWPT}Xiao-Chun Duan, Xiao-Bing Deng, Min-Kang Zhou, Ke Zhang, Wen-Jie Xu, Feng Xiong,
Yao-Yao Xu, Cheng-Gang Shao, Jun Luo and Zhong-Kun Hu, \Journal{\PRL}{117}{2016}{023001}.
\bibitem{FTGKip}K.S. Throne, D.L. Lee and A.P. Lightman, \Journal{\PRD}{7}{1973}{3563}.
\bibitem{WTNEP}Wei-Tou Ni, \Journal{\IJMPD}{25}{2016}{1630002}.
\bibitem{MattGC} V.A. Kosteleck\'y and J.D. Tasson, \Journal{\PRD}{83}{2011}{016013}.
\bibitem{YuriEP}Y. Bonder, E. Fischbach, H. Hernandez-Coronado, D.E. Krause, Z. Rohrbach and D. Sudarsky, \Journal{\PRD}{87}{2013}{125021}.
\bibitem{YuriEPI}Y. Bonder,  \Journal{\PRD}{88}{2013}{105011}.
\bibitem{EXPEPL}M.A. Hohensee, N. Leefer, D. Budker, C. Harabati, V.A. Dzuba and V.V. Flambaum, \Journal{\PRL}{111}{2013}{050401};
V.V. Flambaum, \Journal{\PRL}{117}{2016}{072501}.
%\bibitem{Tino16}G. Rosi, G. D'Amico, L. Cacciapuoti, F. Sorrentino,
M. Prevedelli, M. Zych, C. Brukner, and G. M. Tino, \Journal{\NC}{8}{2017}{15529}.
\bibitem{EntangA}R. Geiger and M. Trupke, \Journal{\PRL}{120}{2018}{043602}.
\bibitem{CVZRW}G. Cognola, L. Vanzo and S. Zerbini, \Journal{\GRG}{18}{1986}{971}.
\bibitem{FW}L.L. Foldy and S.A. Wouthuysen, \Journal{\PR}{78}{1950}{29}.
\bibitem{Case}K.M. Case, \Journal{\PR}{95}{1954}{1323}.

\bibitem{SGI}Y.N. Obukhov,  \Journal{\PRL}{86}{2001}{192}.
\bibitem{FVWMP}H. Feshbach and F. Villars, \Journal{\RMP}{30}{1958}{24}.
\bibitem{ABFW}A. Accioly and H. Blas, \Journal{\PRD}{66}{2002}{067501}.
\bibitem{SilenkoC} A.J. Silenko and O.V. Teryaev, \Journal{\PRD}{71}{2005}{064016}.
\bibitem{WTNH}F.W. Hehl and W.-T. Ni, \Journal{\PRD}{42}{1990}{2045}.
\bibitem{EPIS}E.D. Casola and S. Liberati, \Journal{\AJP}{83}{2015}{39}.
\bibitem{PoiWill} E. Poisson and C.M. Will, Gravity, Newotonian, Post-Newtonian,
 Relativistic, Cambridge University Press, 2014.
\bibitem{QEPMS}H. Hernandez-Coronado and E. Okon, \Journal{\PLA}{377}{2013}{2293}.
\bibitem{SPWEP}L. Seveso and M.G.A. Paris, \Journal{\AP}{380}{2017}{213}.
\bibitem{QEDQG}I.T. Drummond and S.J. Hathrell, \Journal{\PRD}{22}{1980}{343}.
\bibitem{Cartan}E. Cartan {\it C.R.\ Acad.\ Sci.}\ (Paris) {\bf 174}(1922)593.
\bibitem{MAG}Y. Ne'eman and F.W. Hehl, \Journal{\CQG}{14}{1997}{A251};
M. Blagojevi\'c and F.W. Hehl {\it eds.} {\it Gauge Theories of Gravitation}
(London: Imperial College Press, 2013).
\bibitem{QBailey}Q.G. Bailey, V.A. Kosteleck\'y and R. Xu, \Journal{\PRD}{91}{2015}{022006};
B. Altschul {\it et al.}, \Journal{\ASR}{55}{2015}{501}.
\bibitem{ARCL}V.A. Kosteleck\'y and N. Russell, \Journal{\PLB}{693}{2010}{443}.
\bibitem{VCRP} R. Lehnert and R. Potting, \Journal{\PRL}{93}{2004}{110402}; ibid. \Journal{\PRD}{70}{2004}{125010};
B. Altschul, \Journal{\PRL}{98}{2007}{041603}.
\bibitem{VCRCS}C. Kaufhold and F. R. Klinkhamer, \Journal{\NPB}{734}{2006}{1}; ibid. \Journal{\PRD}{76}{2007}{025024}.
\bibitem{VCRF}M. Schreck, \Journal{\PRD}{96}{2017}{095026}.
\bibitem{VCRWZG}D. Colladay, J.P. Noordmans and R. Potting, \Journal{\PRD}{96}{2017}{035034}; ibid.
\Journal{\Sym}{9}{2017}{248}; V.A. Kosteleck\'y and J.D. Tasson,
\Journal{\PLB}{749}{2015}{551}.
%; ibid, \Journal{\JOPC}{873}{2017}{012017}.
\bibitem{LVNRH}V.A. Kosteleck\'y and C. Lane, \Journal{\JMP}{40}{1999}{6245}
\bibitem{bosonA0}M.G. Tarallo, T. Mazzoni, N. Poli, D.V. Sutyrin, X. Zhang and G.M. Tino,
\Journal{\PRL}{113}{2014}{023005};
\bibitem{bosonA}S. Herrmann, H. Dittus and  C. L$\ddot{\mathrm{a}}$mmerzahl, \Journal{\CQG}{29}{2012}{184003}.
\bibitem{ExtGLI}S. Capozziello and M. De Laurentis, \Journal{\PRt}{509}{2011}{167};
A. Delhom-Latorre, G.J. Olmo and M. Ronco, \Journal{\PLB}{780}{2018}{294}.
\bibitem{ExtGLV}V.A. Kosteleck\'y, N. Russell and J.D. Tasson, \Journal{\PRL}{100}{2008}{111102};	
J. Foster, V.A. Kosteleck\'y and Rui Xu, \Journal{\PRD}{95}{2017}{084033};
R. Lehnert, W.M. Snow, Z. Xiao and Rui Xu, \Journal{\PLB}{772}{2017}{865}.
\bibitem{SPINTEST}S.A. Hojman and F.A. Asenjo, \Journal{\CQG}{34}{2017}{ 115011}.
\end{thebibliography}
\end{document}